\documentclass[pra,twocolumn,showpacs,preprintnumbers,amsmath,amssymb]{revtex4}
\usepackage{epsfig}
\usepackage{color}
\usepackage{amsfonts}

\reversemarginpar
\newcommand{\spacer}{\rule[0cm]{0cm}{0cm}}

\newcounter{statementnumber}
\renewcommand{\thestatementnumber}{\arabic{section}.\arabic{statementnumber}}

\newcounter{figurecount}

\def\of{\mathbin{\circ}}

\def\mb{$\begin{displaystyle}}
\def\me{\end{displaystyle}$\ }

\def\Id{\mbox{\rm Id}}

\newcommand{\ket}[1]{\left| #1 \right\rangle}

\newcommand{\twotwo}[4]{\left[ \begin{array}{cc} #1 & #2 \\
                        #3 & #4 \end{array} \right]}

\newcommand{\Stab}{\mbox{\rm Stab}}

\begin{document}

\title{Entanglement verification using local unitary stabilizers}

\author{David W. Lyons}
  \email{lyons@lvc.edu}
\author{Scott N. Walck}
  \email{walck@lvc.edu}
\affiliation{Lebanon Valley College, Annville, PA 17003}

\date{19 June 2013}

\begin{abstract}

\end{abstract}


\begin{abstract}
Local unitary stabilizer subgroups constitute powerful invariants for
distinguishing various types of multipartite entanglement. In this
paper, we show how stabilizers can be used as a basis for entanglement
verification protocols on distributed quantum networks using minimal
resources. As an example, we develop and analyze the performance of a
protocol to verify membership in the space of Werner states, that is,
multi-qubit states that are invariant under the action of any 1-qubit
unitary applied to all the qubits.
\end{abstract}

\maketitle

\section{Introduction}

Entangled multipartite quantum states of two-level quantum systems, or
qubits, play a key role as resources used in quantum computation and
quantum communication protocols (see, for example,
\cite{bennett93,shor94}, or \cite{nielsenchuang,gudder03} for more
more extensive surveys). In this paper, we
consider a basic practical problem in distributed quantum computing: a
collection of $n$ parties will perform a distributed quantum
computational task that requires as a resource a particular entangled
state, or more generally, any state from a given family of entangled
states. An untrusted source provides the resource states, and the
parties wish to verify that a resource state is genuine. To minimize the
cost of verification, the parties have access to only a limited
collection of single-qubit operations. For a survey of entanglement
verification protocols, see~\cite{vanenk2007}, and for recent work
related to this paper, see~\cite{pappa2012} and its references.

Inspired by \cite{pappa2012}, in which Pappa et al. treat the
verification test problem for the $n$-particle GHZ state, we present
here a general framework for a verification test protocol for certain
types of families of states, namely, families characterized by
stabilizing subgroups of the local unitary group. This is motivated by
the authors' previous
work~\cite{maxstabnonprod2,symmstates2,symmmixed,wernerstructure} that
demonstrates a precise connection between many well-known entanglement
resources, such as the GHZ states, Werner states, and permutationally
invariant states that include the $W$ states and Dicke states, and their
stabilizer subgroups of the local unitary group.

To illustrate the general framework of the subgroup-based verification
protocol, we give a specific test for verification of the family of
Werner states, which are those states stabilized by the subgroup of
local unitary operators that consist of the same 1-qubit unitary acting
on all $n$ qubits.

The paper is organized as follows. We begin in Section~\ref{stabinv}
with the relationship between states and subgroups of the local unitary
group. Section~\ref{genframe} gives the general framework for
entanglement verification tests that exploit the connection of states
with subgroups. Then in Section~\ref{wernertest} we give an example of
how the general test framework can be used in the case of the family of
Werner states.

\section{The local unitary stabilizer as entanglement invariant}\label{stabinv}

The local unitary (LU) group $G$ on $n$-qubit states is the group of
1-qubit operations. $G$ can be taken to be
$G=U(1)\times SU(2)^n$ in the case of pure states, or more simply
$G=SU(2)^n$ for mixed states. An element $g\in G$ acts via
multiplication by the Kronecker product of its components. Explicitly,
for pure states, the element $g=(e^{it},g_1,\ldots,g_n)\in
G$ acts on $\ket{\psi}$ by 
$$g\ket{\psi}:=e^{it}(g_1\otimes \cdots\otimes
g_n)\ket{\psi}.$$

Given a pure state $\ket{\psi}$ or a mixed state $\rho$ of $n$-qubits,
the stabilizer subgroup of the local unitary group is defined by
\begin{eqnarray*}
  \Stab_\psi &=&\{g\in G\colon g\ket{\psi}=\ket{\psi}\}\\
  \Stab_\rho &=&\{g\in G\colon g\rho g^\dagger=\rho\}
\end{eqnarray*}
Two basic facts that make stabilizer
groups useful in the study of entanglement are as follows.
\begin{enumerate}
\item If two states are LU-equivalent, then their stabilizers are
  isomorphic. 
\item Many classes of known useful entanglement resources are
  characterized by their stabilizer group.
\end{enumerate}
A consequence of the first statement is that the isomorphism class (more
precisely, the conjugacy class) of a stabilizer subgroup is an
LU-invariant of the corresponding state~\cite{symmstatespaper}. The
second statement summarizes a program carried out in a series of papers
for pure and mixed Werner states and permutationally invariant
states~\cite{maxstabnonprod2,symmstates2,symmmixed,wernerstructure}.

The basic idea used in the verification test framework in the next
section is the following. We wish to verify whether a given state
$\ket{\Psi}$ is a member of a family $V$ of entanglement resource states
that are stabilized by all local unitary operations in a subgroup $S$ of
the local unitary group. We have at our disposal a measurement $M$ and a
function $F$ so that $F\of M$ is constant on $V$.  It may be that there
are counterfeit states, that is, states not in $V$, which nonetheless
pass our measurement test by yielding the same result as 
$F\of M$ applied to elements of $V$. Here is where the stabilizing group
$S$ comes in. If $\ket{\Psi}$ is not a genuine member of $V$, there are
local unitary operations in $S$ that take $\ket{\Psi}$ to a new state
$\ket{\Psi'}\neq \ket{\Psi}$. Our strategy is to carefully choose a
small subset of elements of $T\subset S$ and the function $F$ so that by
randomly applying elements of $T$, we guarantee a nonzero probability
that the composition $F\of M$ applied to $\ket{\Psi'}$ yields a result
different from the value of $F\of M$ on $V$, and thus will detect
counterfeit states.

\section{Entanglement verification using local unitary stabilizers}\label{genframe}

Given a subgroup $S$ of the local unitary group, let
$V(S)$ denote the space of states fixed by every element of $S$, that is, 
$$ V(S)= \{\ket{\psi}\colon g\ket{\psi}=\ket{\psi} \mbox{ for all } g\in S\}.
$$

{\bf The verification task.} An untrusted source produces $n$-qubit pure
states that are claimed to be members of a subspace $V=V(S)$ for some
subgroup $S$ of the local unitary group. The source
distributes the qubits of each state it produces, one qubit to each of
$n$ parties. One party, the verifier, seeks to determine whether the
states being produced are members of $V$. Each of the parties possesses
a trusted 1-qubit measurement device that measures in the standard
basis, and has the ability to apply any of a finite collection of
1-qubit gates to their own qubit. Each party shares a trusted classical
communication channel with the verifier.

{\bf Set-up for the verification test.} The verifier chooses a
probability distribution $p_1,p_2,\ldots,p_k$ and elements
$g_1,g_2,\ldots,g_k$ in $S$ and a function $F$ that takes as inputs
Boolean strings of length $n$ (these come from measuring states in the
standard computational basis) and produces integer or Boolean
outputs. $F$ is required to be well-defined on $V$ in the sense that if
we measure a state $\ket{\psi}\in V$ in the standard basis, then apply
$F$ to the result, then we must get the same output (which, without loss
of generality, we may take to be zero), no matter what
post-measurement state was obtained. To be precise, this means that if
$$\ket{\psi}=\sum_I c_I \ket{I}$$ is the expansion of $\ket{\psi}$ in
the standard basis (here, $I=i_1i_2\cdots i_n$ denotes an $n$-bit string, with $i_k=0,1$
for $1\leq k\leq n$) then we must have $F(I)=0$ whenever
$c_I\neq 0$.

The verifier chooses elements $g_i$ and
the function $F$. Let 
$$V_0=\{\sum c_I\ket{I}\colon c_I\neq 0 \Rightarrow
F(I)=0\} $$
denote the space
of states $\ket{\phi}$ for which $F(x)=0$ for every possible value
$x$ obtained from measuring $\ket{\phi}$ in the standard basis.
The verifier desires the following two
properties. 
\begin{enumerate}
\item [(i)] $F(x)=0$ for every $x$ obtained by measuring states in $V$
  (in other words, $V\subseteq V_0$), and
\item [(ii)] for every $\ket{\psi}$ in $V_0\setminus V$, there is a $g_i$
  such that $g_i\ket{\psi}$ is {\em not} in $V_0$.
\end{enumerate}
Whether it is possible to achieve both of these properties will depend upon
the local unitary subgroup in question.

Each $g_i$ is of the form $g_i=U^i_1\otimes U^i_2\otimes \cdots \otimes
U^i_n$ for some $U^i_j$ in $U(2)$. Each party $j$ is provided with the
ability to execute the 1-qubit gates $U^1_j, U^2_j,\ldots,U^k_j$.

{\bf Protocol for the verification test.} Given a state $\ket{\Psi}$
produced by the source, with qubits distributed to the $n$ parties, the
verifier randomly selects one of the $g_i$ with probability $p_i$. The
verifier instructs each party $j$ to apply $U^i_j$ to their qubit, so
that the state is now $g_i\ket{\Psi}$. Then each party measures their
qubit in the computational basis and reports the result to the
verifier. The verifier applies $F$ to the binary string $x$ resulting from the
measurements. The verifier accepts $\ket{\Psi}$ if $F(x)=0$, and rejects
otherwise. 

It is clear that the protocol accepts states in $V$ with probability
1. It is less obvious but intuitively plausible that the protocol
accepts a counterfeit state $\ket{\Psi}\not\in V$ with probability
bounded below 1. Here is the heuristic argument. If $\ket{\Psi}\not\in
V$, there is some $g_i$ such that $g_i\ket{\Psi}\not\in V_0$.  There is
a nonzero probability $p_i$ that the verifier chooses $g_i$ in the first
step of the protocol. In this case, when we now measure $g_i\ket{\Psi}$
and apply $F$, we are {\em not guaranteed} to get 0, and it seems likely
that there should be a way to engineer the $g_i$s so that there is a
guaranteed nonzero probability that we get an $F$ value
other than zero. The difficult part of proving the effectiveness of a
protocol is showing that the collection of $g_i$s and the function $F$
do indeed guarantee a nonzero probability of rejecting
$\ket{\Psi}\not\in V$. In the next section, we give the details and
prove that the protocol works for a test to verify membership in the
Werner subspace.

\section{Example: Test for pure Werner subspace membership}\label{wernertest}

The space ${\cal W}$ of $n$-qubit pure Werner states is defined to be those states $\ket{\psi}$
that satisfy
$$ U^{\otimes n} \ket{\psi} = \ket{\psi}
$$
(up to phase) for all $2\times 2$ unitaries $U$. The space ${\cal W}$ is
sometimes called the {\em decoherence free subspace for collective
  decoherence}. See the introduction to~\cite{wernerstructure} for a brief survey of the
uses and importance of this class of states.

It is known~\cite{su2blockstates}
that there are no pure Werner states for odd $n$ qubits, and the space of
pure Werner states for even $n$ qubits is spanned by the set of products of
singlet states $\frac{1}{\sqrt{2}}(\ket{01}-\ket{10})$ in various pairs of qubits.


{\bf Set-up for the Werner verification test.} An untrusted source produces an $n$-qubit
state $\ket{\Psi}$, and distributes the qubits, one to each of $n$
parties. The verifier, say, Party 1, chooses probability distribution
$p_0=p_1=1/2$ and local unitaries $g_0=\Id,g_1=H^{\otimes n}$, where $H$
is the Hadamard matrix. Let $F$ be the function that returns the number
of 1's minus the number of 0's in a $n$-bit Boolean string. Observe that
the value of $F$ is zero on all strings $x$ that result from measuring a
Werner state in the standard basis.

{\bf Werner verification protocol.} The verifier selects either $g_0,g_1$ with equal
probability, and instructs all parties to apply a Hadamard gate to their
qubit if $g_1$ is selected. 
Each party $j$ reports measurement result $x_j$. The verifier evaluates
$F(x)=F(x_1x_2\cdots x_n)$ and accepts $\ket{\Psi}$ if the the result is
0, and rejects otherwise.

Because any Werner state $\ket{\Psi}$ is a superposition of singlet
products, the protocol clearly accepts $\ket{\Psi}$ when $g_0$ is selected. Because
$H^{\otimes n}\ket{\Psi}=\ket{\Psi}$ and $H$ takes the $\ket{0},\ket{1}$
basis to the $\ket{+},\ket{-}$ basis, the protocol accepts $\ket{\Psi}$
when $g_1$ is selected. The following theorem shows that the probability
of false acceptance of a counterfeit state is bounded by a probability
less than 1.

{\bf Theorem 1.} Suppose that $\langle \Psi | P_{\cal W} | \Psi \rangle =
1-\epsilon^2$, where $P_{\cal W}$ is projection onto the Werner subspace. Then
we have 
$${\rm Pr}(\mbox{accept }\Psi) \leq 1-\frac{\epsilon^2}{2}(1-m)
$$
where $m<1$ is the maximum fidelity $\max \{|\langle {u}|{v}\rangle |\}$
between normalized states
$\ket{u},\ket{v}$ that lie in subspaces to be described below.

{\bf Comments.} 
We point out that our protocol recovers
Theorem~1 of~\cite{pappa2012} for $n=2$. In that case, the Werner
subspace is the 1-dimensional span of the singlet, which is local
unitary equivalent to $\ket{\Phi^2_0}$ in~\cite{pappa2012}. The maximum
inner product $m$ in our Theorem~1 is $0$, so the probability of acceptance is the
same as in~\cite{pappa2012}.

{\bf Proof of Theorem 1.} It is convenient to give names to a number of
subspaces of the Hilbert space ${\cal H}$ of pure states of $n=2k$
qubits. As we have already noted, let ${\cal W}$ denote the Werner
subspace. Let $S$ be the span of the weight $k$ standard basis vectors,
and let $T$ be the span of the weight $k$ $+,-$ basis vectors. A key
fact that makes Theorem~1 possible is that ${\cal W}=S\cap T$ (see
Lemma~1 in the Appendix).  Let $U={\cal W}^\perp \cap S$ and let $V={\cal W}^\perp \cap
T$. Let $S'=S^\perp \cap (S+T)$ and let $T'=T^\perp \cap
(S+T)$. Finally, let $L=(S+T)^\perp$. For each of these spaces $A$,
let $P_A$ denote the projection onto $A$.  Let $m$ denote the maximum 
inner product between unit vectors in $U,V$, that is,
$$ m=\max\{|\langle \psi|\phi\rangle| \colon \ket{\psi}\in U,
\ket{\phi}\in V\}, \langle\psi|\psi\rangle = \langle\phi|\phi\rangle = 1\}.
$$

We can write any $\ket{\Psi}$
in ${\cal H}$ as an orthogonal sum 
$$ \ket{\Psi} = P_{\cal W}\ket{\Psi} + P_{U+V}\ket{\Psi} + P_{L}\ket{\Psi}.
$$
We further decompose $P_{U+V}\ket{\Psi}$ in two different
orthogonal sums.
\begin{eqnarray}
  P_{U+V}\ket{\Psi} &=& P_U\ket{\Psi} + P_{S'}\ket{\Psi}\\
  P_{U+V}\ket{\Psi} &=& P_V\ket{\Psi} + P_{T'}\ket{\Psi}
\end{eqnarray}
Given that the trace distance from $\ket{\Psi}$ to the nearest pure
Werner state vector is $\epsilon$, that is, $\langle \Psi | P_{\cal W} |
\Psi \rangle = 1-\epsilon^2$, we can define
$\epsilon_1,\epsilon_2,\alpha,\beta$ (positive quantities that sum to
$\epsilon^2$) as follows.
\begin{eqnarray}
\langle \Psi | P_{U+V} | \Psi \rangle &=& \epsilon_1^2\\
\langle \Psi | P_{U} | \Psi \rangle &=& \alpha^2\\
\langle \Psi | P_{V} | \Psi \rangle &=& \beta^2
\end{eqnarray}
The probability of acceptance of $\ket{\Psi}$ by our protocol is given
by 
\begin{eqnarray*}
{\rm Pr}(\mbox{accept }\Psi) &=& \frac{1}{2}\langle \Psi | P_{S} | \Psi
\rangle + \frac{1}{2}\langle \Psi | P_{T} | \Psi \rangle\\
&=& 1-\epsilon^2 + \frac{\alpha^2+\beta^2}{2}.
\end{eqnarray*}
Applying Lemma~2 in the Appendix to the unit vector
$\frac{P_{U+V}\ket{\Psi}}{\langle \Psi | P_{U+V} | \Psi \rangle}$, we
have $\alpha^2 + \beta^2\leq (1+m)\epsilon_1^2$, which is in turn less
than or equal to $(1+m)\epsilon^2$. Thus we conclude that the inequality in the statement of
Theorem~1 holds.

Finally, we argue that $m$ is guaranteed to be less than 1. It suffices
to observe that (i) $U\cap V=\{0\}$, and (ii), neither $U$ nor $V$ is
the zero subspace. Condition (i) is guaranteed by construction:
$U\subset S$ and $V\subset T$ are orthogonal complements in $S,T$,
respectively, to the
intersection ${\cal W}=S\cap T$ (see Lemma~1 in the Appendix for the
proof that the Werner subspace ${\cal W}$ is indeed the intersection
$S\cap T$). Condition (ii) follows from considering dimensions. For
$n=2k$, we have
$\dim {\cal W} = \frac{1}{k+1}{2k \choose k}$ (see, for example,
\cite{su2blockstates}), which is less than $\dim S= \dim T= {2k\choose
  k}$.  

This concludes the proof of Theorem~1.

\section{Outlook}
A natural next question for our stabilizer based testing protocol
framework is quantifying performance. For example, in the case of the
Werner test protocol, it is natural to ask about the value of $m$
(beyond simply $m<1$) and the optimality of the bound in Theorem~1. For
$n=2$, everything is explicit. We have
\begin{eqnarray*}
S &=& \mbox{span of } \ket{01},\ket{10}\\
T &=& \mbox{span of } \ket{+-},\ket{-+}\\
{\cal W} &=& \mbox{span of } \ket{01}-\ket{10}\\
U &=& \mbox{span of } \ket{01}+\ket{10}\\
V &=& \mbox{span of } \ket{+-}+\ket{-+}
\end{eqnarray*}
and it is clear that the maximum inner product $m$ is 0.

For higher numbers of qubits, one can find $m$ as follows. It is a
simple exercise to verify that, for subspaces $U,V$ with orthonormal bases $\ket{i},\ket{j}$,
the maximum overlap 
$$ m=\max\{|\langle \psi|\phi\rangle| \colon \ket{\psi}\in U,
\ket{\phi}\in V\}, \langle\psi|\psi\rangle = \langle\phi|\phi\rangle = 1\}.
$$ is the largest singular value of the matrix whose $i,j$ entry is
$\langle i | j\rangle$.  Extending this slightly, we have that $m$ is
the largest singular value {\em less than 1} of the matrix whose $i,j$
entry is $\langle i | j\rangle$, where $\ket{i},\ket{j}$ are orthonormal
bases for $S$ and $T$. Computer-assisted numerical calculations show
that $m=1/2$ for $n=4,6,8,10$. We conjecture that $m=1/2$ for $n=2k$
with $k\geq 2$, and further, that the bound in Theorem~1 is optimal.

Another avenue of investigation is to apply the general stabilizer framework to
further classes of states.

{\bf Acknowledgments.} The authors thank Elena Diamanti and Damian Markham for
helpful conversations, and also Philip Spain for pointing out a useful
reference for the proof of Lemma~2. We thank the anonymous referee for a
thoughtful review that led to several improvements. This work was
supported by NSF Award No. PHY-1211594.

\bibliographystyle{unsrt}



\section{Appendix}

{\bf Lemma 1.} The Werner subspace ${\cal W}$ of $n=2k$ qubit pure
states, that is,
$${\cal W}=\{\ket{\psi} \colon U^{\otimes n} \ket{\psi}=\ket{\psi}
\mbox{ for all } U\in U(2)\},
$$
is the intersection $S\cap T$ of the subspaces 
\begin{eqnarray*}
  S &=& \mbox{span of the weight $k$ basis vectors in the $0,1$
  basis}\\
  T &=& \mbox{span of the weight $k$ basis vectors in the $+,-$ basis}.
\end{eqnarray*}

{\bf Proof.} We show in~\cite{su2blockstates} that any pure Werner state is a
superposition of products of the singlet state
$\frac{1}{\sqrt{2}}(\ket{01}-\ket{10})$, so it is clear that ${\cal
  W}\subset S$. The Hadamard matrix
$H=\frac{1}{\sqrt{2}}\twotwo{1}{1}{1}{-1}$ takes $\ket{0},\ket{1}$ to
$\ket{+},\ket{-}$, so $H^{\otimes n}$ takes $S$ to $T$. Since
$H^{\otimes n}$ fixes every state in ${\cal W}$, we have ${\cal
  W}\subset T$. Conversely, suppose we have a state $\ket{\psi}$ in
$S\cap T$. Because $\ket{\psi}\in S$, it is
a simple calculation that 
$$\left.\frac{d}{dt}\right|_{t=0}\left(\exp(itZ)\otimes\cdots\otimes\exp(itZ)\right)\ket{\psi}=0$$
where $Z$ is the Pauli $Z$ matrix (see, for example,~\cite{maxstabnonprod2}). Likewise, because $\ket{\psi}\in
T$, we have
$$\left.\frac{d}{dt}\right|_{t=0}\left(\exp(itX)\otimes\cdots\otimes\exp(itX)\right)\ket{\psi}=0$$
where $X$ is the Pauli $X$ matrix. Because $(iZ,iZ,\ldots,iZ)$,
$(iX,iX,\ldots,iX)$ generate the Lie algebra of the subgroup $\Delta =
\{(U,U,\ldots,U)\colon U\in SU(2)\}$ that defines the Werner subspace,
we conclude that $\ket{\psi}$ is in ${\cal W}$.

{\bf Lemma 2.} Let $U,V$ be subspaces of a Hilbert
space $H$ of any dimension, and let $P_U,P_V$ be the orthogonal projections onto
$U,V$, respectively. Let $\ket{\Psi}$ be a unit vector $H$, and let 
$$ m=\max\{|\langle \psi|\phi\rangle| \colon \ket{\psi}\in U,
\ket{\phi}\in V\}, \langle\psi|\psi\rangle = \langle\phi|\phi\rangle = 1\}.
$$
Then 
$$ \langle \Psi|P_U|\Psi\rangle + \langle \Psi|P_V|\Psi\rangle \leq m+1.
$$

{\bf Proof.} Let $\ket{\Psi}$ be given. Define unit vectors
$\ket{u}\in U, \ket{v}\in V$ 
by 
\begin{eqnarray*}
  \ket{u} &=& \frac{P_U\ket{\Psi}}{\sqrt{\langle \Psi | P_U |\Psi\rangle}}\\
  \ket{v} &=& \frac{P_V\ket{\Psi}}{\sqrt{\langle \Psi | P_V |\Psi\rangle}}
\end{eqnarray*}
(if $P_U\ket{\Psi}=0$, choose any unit vector in $U$ for $\ket{u}$,
and similarly for $\ket{v}$ in the case that $P_V\ket{\Psi}=0$)
so that we have
\begin{eqnarray*}
  \langle \Psi | P_U |\Psi\rangle &=& |\langle \Psi|u\rangle |^2\\
  \langle \Psi | P_V |\Psi\rangle &=& |\langle \Psi|v\rangle |^2.
\end{eqnarray*}
A generalization of Bessel's inequality~\cite{boas1941} says that
$$ |\langle \Psi|u\rangle|^2 + |\langle \Psi|v\rangle|^2 \leq 1 + |\langle u|v\rangle|
$$
so we have
\begin{eqnarray*}
&&  \langle \Psi | P_U |\Psi\rangle + \langle \Psi | P_V |\Psi\rangle\\
 &=& |\langle \Psi|u\rangle |^2 + |\langle \Psi|v\rangle |^2\\
&\leq& 1 + |\langle u|v\rangle|\\
&\leq & 1+m.
\end{eqnarray*}

\end{document}